# Photonic band structure and eigenmodes of magnetophotonic crystals

A. B. Khanikaev,[1,2] K. Yayoi,[3] M. J. Steel,[2] A. V. Baryshev,[1] and M. Inoue[1]

[1] *Toyohashi University of Technology, 1-1, Tempaku, Toyohashi, Aichi 441-8580, Japan*

[2] *MQ Photonics Research Centre and CUDOS, Department of Physics, Macquarie University, NSW 2109, Australia*

[3] *Ibaraki National College of Technology, 866, Nakane, Hitachinaka, Ibaraki 312-8508, Japan*

*Corresponding author: khanikaev@gmail.com*

We study photonic band structure of two- and three-dimensional magnetophotonic crystals and the polarization properties of their eigenmodes using a plane wave expansion method. The alteration of the photonic band structure and eigenmodes by magnetization are examined. Two orientations are studied: in-plane magnetization and perpendicular magnetization. The magnetization-induced symmetry breaking effects are shown to be responsible for the formation of exotic magnetic Bloch modes, comprised of two or more coupled symmetric and antisymmetric Bloch modes of the corresponding non-magnetic photonic crystal. These results imply that uncoupled states of non-magnetic photonic crystals can be excited, or become coupled, in magnetic structures. We show that the polarization state of magnetic Bloch modes is very complicated. In the particular case of perpendicular to plane magnetization they represent in-plane elliptically (or even circularly) polarized waves with large longitudinal component.

1. Introduction

Magnetophotonic crystals (MPCs) represent a class of artificial magneto-optic (MO) materials, having a periodically modulated structure, with at least one of the constituent materials possessing MO activity. Unique optical and MO properties of MPCs stem from the subtle interplay of MO activity and Bragg diffraction, which enables dynamic tuning of the photonic crystal eigenmodes and photonic band structure (PBS) by controlling the magnetization with an external applied magnetic field [1,2,3,4,5,6]. Huge enhancement of the magneto-optic effects, due to the low group velocity of Bloch modes [1,5,7] near band edges of the photonic crystals (PCs) or due to strong localization of the electromagnetic radiation at the defects introduced in photonic crystals have also been demonstrated [1,8-10]. The huge field enhancement of a defect mode when resonantly coupled to waveguides raises the possibility of construction of multiport devices with controllable transport properties such as isolators and circulators. [12].

Nevertheless, even though MPCs were proposed almost ten years ago, there is a lack of understanding of their general optical and MO properties. Most early works dealing with MPCs were either devoted to study of one dimensional structures [8,9] or quasi-one-dimensional models [5], which can be treated analytically by transfer matrix technique, or utilize approximate perturbation approaches [13,14]. However, quite recently several new extraordinary phenomena were predicted theoretically. Edge states similar to those found in 2D quantum Hall systems [15,16] and suppression of backscattering for unidirectional modes of MPCs [17,18] demonstrate that the potential impact of magneto-optical activity in 2D and 3D periodic systems remains rather unexplored.

In the present work we aim at least partially, to fill in the gap in understanding of the basic properties of 2D and 3D magneto-photonic crystals. We apply a generalization of the plane wave expansion method and calculate the PBS and eigenmodes of MPCs in different magnetization geometries and analyze their polarization state.

The plane wave expansion method for the MPCs is briefly considered in Sec. 2. The calculation results for the 2D square and 3D simple cubic lattices for different magnetization geometries are presented and discussed in Sec. 3. The conclusions of the work are summarized in Sec. 4.

**2. Generalized plane wave method**

The plane wave expansion (PWE) method [19] has been proven to be the one of the most convenient and powerful methods for study of the PBS and electromagnetic eigenmodes of infinite PCs. Moreover, the knowledge of the PBS and eigenmodes of infinite structures allows one to study the optical properties and evaluate MO response of finite structures by mode-matching techniques [20]. Hence, knowledge of the band structure along with polarization state of the eigenmodes is central to understanding the optical and MO response of real MPC. These arguments make the PWE method the first tool which can be applied in studies of harmonic wave propagation in periodical dielectric MO media. For the sake of completeness, in this section we derive a generalized PWE method for the electric field, which can be applied to the case of PCs consisting of MO media.

Our starting point is the wave equation in MO media. Using the gyration vector notation (as opposed to a tensor permittivity), the wave equation takes the form [21]:

$$\frac{1}{\varepsilon(\mathbf{r})}\left[\nabla\times[\nabla\times\mathbf{E}(\mathbf{r},t)]\right] = -\frac{1}{c^2}\frac{\partial^2}{\partial t^2}\left(\mathbf{E}(\mathbf{r},t) + i[\mathbf{E}(\mathbf{r},t)\times\mathbf{g}(\mathbf{r})]\right), \quad (1)$$

where as usual $\mathbf{E}(\mathbf{r},t)$ is the electric field, $\varepsilon(\mathbf{r})$ is the scalar permittivity, $\mathbf{g}(\mathbf{r})$ is the gyration vector, and $c$ is the light velocity in vacuum. We consider propagation of electromagnetic waves in a periodically modulated structure and assume the gyration vector and permittivity share the same periodicity:

$$\varepsilon(\mathbf{r}) = \varepsilon(\mathbf{r}+\mathbf{a}_i), \ \mathbf{g}(\mathbf{r}) = \mathbf{g}(\mathbf{r}+\mathbf{a}_i), \quad (2)$$

where $\mathbf{a}_i$ ($i$=1,2,3) are the elementary vectors of the lattice. Spatial periodicity of the MPC allows us to expand the inverse permittivity $\varepsilon(\mathbf{r})^{-1}$ and gyration vector $\mathbf{g}(\mathbf{r})$ in the Fourier series:

$$\varepsilon(\mathbf{r})^{-1} = \sum_{\mathbf{G}} \kappa_{\mathbf{G}} \exp\{i\mathbf{G}\cdot\mathbf{r}\}, \quad (3)$$

$$\mathbf{g}(\mathbf{r}) = \hat{\mathbf{m}}\sum_{\mathbf{G}} \eta_{\mathbf{G}} \exp\{i\mathbf{G}\cdot\mathbf{r}\}, \quad (4)$$

where $\hat{\mathbf{m}}$ is a unit vector in the direction of the magnetization (assumed fixed) and $\mathbf{G}$ ranges over the reciprocal lattice vectors. Then, using Bloch's theorem for the field expansion in the form

$$\mathbf{E}(\mathbf{r}) = \sum_{\mathbf{G}} \mathbf{c}_{\mathbf{G}}^{kn} \exp\{i(\mathbf{k}+\mathbf{G})\cdot\mathbf{r}\}, \quad (8)$$

and substituting (3), (4) and (5) into (1), we obtain an eigenvalue equation for the expansion coefficients $\mathbf{c}_{\mathbf{G}}^{kn}$:

$$-\sum_{\mathbf{G}'} \kappa_{\mathbf{G}-\mathbf{G}'}(\mathbf{k}+\mathbf{G}')\times\left\{(\mathbf{k}+\mathbf{G}')\times\mathbf{c}_{\mathbf{G}'}^{kn}\right\} = \frac{\omega_{kn}^2}{c^2}\left(\mathbf{c}_{\mathbf{G}}^{kn} + i\sum_{\mathbf{G}'}\eta_{\mathbf{G}-\mathbf{G}'}\left\{\mathbf{m}\times\mathbf{c}_{\mathbf{G}'}^{kn}\right\}\right). \quad (9)$$

Here $\omega_{\mathbf{k}n}$ denotes the eigenfrequency corresponding to the eigenvector $\mathbf{c}_{\mathbf{G}}^{\mathbf{k}n}$ for a particular Bloch vector $\mathbf{k}$ and band $n$. We label the corresponding spatial Bloch mode by $\mathbf{E}_{\mathbf{k}n}$. By solving (9) numerically, in addition to the PBS one is able to extract eigenmodes of an MPC and study their polarization state and evaluate MO properties of the finite structures.

The system (9) represents a generalized eigenvalue problem [22] of the form $\hat{A}\mathbf{c}^{\mathbf{k}n} = \omega_{\mathbf{k}n}\hat{B}\mathbf{c}^{\mathbf{k}n}$, which for PCs formed from non-magnetic materials simplifies to the standard eigenvalue problem $\hat{A}\mathbf{c}^{\mathbf{k}n} = \omega_{\mathbf{k}n}\mathbf{c}^{\mathbf{k}n}$, well-known for conventional PCs (in this case matrix $\hat{B}$ reduces to the identity matrix). Therefore, one can consider a non-magnetic PC as a particular case of a magnetic one for which the matrix $\hat{B}$ is the identity matrix. This form of the eigenvalue problem makes the physical consequences of the MO activity on eigenmodes very transparent. The non-vanishing of the off-diagonal elements introduces conversion or mixing [5] of the different eigenmodes of the non-magnetic PC. However, because of the natural weakness of the MO activity these non-diagonal elements are small in comparison with the unities along the diagonal of $\hat{B}$, and therefore in most cases, the mode mixing is weak and direct correspondence between solutions of the non-magnetic and magnetic problems can be ascertained. As will be shown below, exceptions to this rule are degeneracies where alteration of the band structure and eigenmodes is most profound.

Note that eigenvalue problem (9) represents only one of several possible formulations. A conventional eigenvalue problem can be formulated for the magnetic field $\mathbf{H}$ as well [2] in the more familiar form

$$\nabla \times \hat{\varepsilon}^{-1} \nabla \times \mathbf{H} = \frac{\omega^2}{c^2}\mathbf{H}, \tag{10}$$

where $\hat{\varepsilon}$ is the permittivity tensor including off-diagonal magnetic terms. While being less time consuming and demanding less numerical resources, this does not provide such a simple physical picture as the generalized problem (9) since the magnetic effects are embedded inside the permittivity and their impact as a perturbation is less apparent. In addition, a polarization analysis can be straightforwardly conducted with the use of $\mathbf{E}$-eigenvectors calculated from (9), while knowledge of the $\mathbf{H}$-eigenvectors requires additional calculations. All calculations presented below were conducted for both the $\mathbf{E}$ and $\mathbf{H}$ problems and no discrepancies in calculation results were observed, while the convergence speed was approximately the same for both formulations.

## 3. Two-dimensional magnetophotonic crystals

It is well known that due to a mirror symmetry of 2D PCs, the vector eigenvalue problem (9) for conventional non-magnetic PCs can be reduced to two independent scalar eigenvalue problems giving rise to the existence of two types of modes—TE and TM modes with the electric field vector $\mathbf{E}_{kn}$ parallel or perpendicular to the plane of the 2D structure, respectively [19]. However, when materials constituting the structure possess MO activity, the additional term on the right-hand side of the vector wave equation (1) or the matrix $\hat{B}$ on the right-hand side in the eigenvalue problem (9) couples these modes. Except for the case of magnetization perpendicular to the plane, the separation of modes is impossible and one has to solve a fully vectorial generalized eigenvalue problem.

Nevertheless, solutions of the generalized problem (9) still share some features with solutions of the common eigenvalue problem for the non-magnetic PCs. As was mentioned before, existence of the non-diagonal elements in the matrix $\hat{B}$ of (9) introduces conversion or mixing of the TM and TE modes independent in the non-magnetic PCs. However, because of the weak character of the magneto-optic activity these non-diagonal elements are small, in comparison with unity at the diagonal, the mode mixing is weak and direct correspondence between solutions of the nonmagnetic (pure TE and TM modes) and magnetic (mixed TE and TM modes) problems can be easily ascertained. The exception is in the vicinity of the high symmetry points of the Brillouin zone where TE and TM modes are mutually degenerate as well as near points of accidental degeneracy. Analysis based on the perturbation theory shows that in such cases, even infinitesimally weak magneto-optic activity leads to behavior similar to hybridization effects in electronic systems and causes strong coupling of the TE and TM modes and substantial alteration of the PBS [2,13,14]. In the next two sections we consider the effects of the two orientations of the field in turn.

### 3.1 Two-dimensional magnetophotonic crystals magnetized in-plane

We begin with the case of magnetization in the plane, for which the TE/TM separation is not possible. We consider the PBS for a MPC of circular magnetic rods in air arranged in a square lattice and magnetized along the $\Gamma X$-direction. The rods have radius $r=0.4$, permittivity $\varepsilon=6.25$ and magnitude of the gyration vector $|\mathbf{g}|=0.03$, corresponding to the Cerium-substituted Yttrium Iron Garnet (Ce:YIG) at optical communication wavelength. $\lambda_0=1550$ nm [11,12]. Figure 1 shows the band structure for both the magnetic structure (black lines) and for TE (blue) and TM (red) modes of the corresponding non-magnetic structure with $\mathbf{g}=0$. At this scale, the difference in calculated PBS for the magnetic and non-magnetic cases is not visible. To make this difference more apparent, we enlarge regions around the high symmetry $\Gamma$ and M points (Figs 2(a) and 3(a), respectively) where triple degeneracy takes place, as well as in the proximity of an accidental degeneracy between the fourth TE and TM bands (Fig. 4(a)). At the $\Gamma$ point in Fig. 2(a), the one TE and two TM modes are mutually degenerate. As the Bloch vector moves away from the symmetry point (i.e. to the right hand part of the figure), the frequencies of the hybridized magnetic modes tend back to the non-magnetic branches. Similar behavior is seen at the M point in Fig. 3(a). As expected from our preliminary consideration, the alteration of the PBS is more pronounced at the degeneracies between TE and TM polarized modes of non-magnetic PC. When moving apart from the degeneracies, the effect of magnetization gradually tails off and the branches asymptotically tend to those of nonmagnetic PCs. Following a tradition to compare PCs with semiconductors, the magnetization-induced splitting of the dispersion curves and the removal of degeneracy for bands of different light helicity can be considered as an optical counterpart of the Zeeman Effect [5].

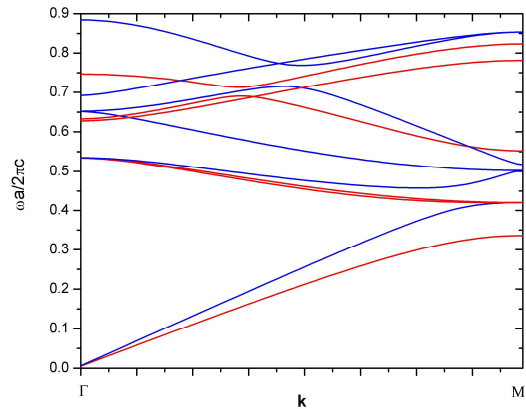

Fig. 1. (Color online) Photonic band structure of 2D photonic crystal comprised of square lattice of circular magneto-optical rods (black lines), and TE (blue) and TM (red) modes of the non-magnetic lattice. Physical parameters are given in the text.

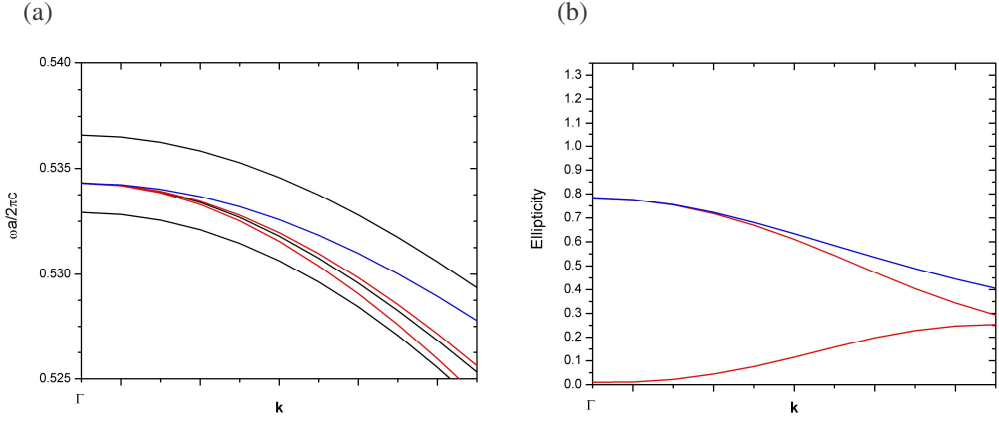

Fig. 2. (Color online) (a) Photonic band structure of MPC (shown by black lines) and non-magnetic PC (shown by blue and red lines for TE and TM polarized modes, respectively) in the proximity of Γ point, and (b) associated ellipticity of hybrid Bloch eigenmodes of MPC; The different line styles for the magnetic modes indicate correspondences between the curves in (a) and (b).

*3.1.1 Eigenmode hybridization*

While the frequency shifts are weak, a significantly more pronounced effect of the MO activity on the eigenmodes can be seen in the change of their polarization state in the proximity of the degeneracies. The common notion of ellipticity is not applicable for PCs, since every Bloch mode is formed from a number of plane waves each having their own polarization characteristics [23]. We can, however, introduce a generalization of ellipticity or mode hybridization which reduces to the standard ellipticity in the long wavelength limit. To do this we rewrite the field expansion coefficients $\mathbf{c}_\mathbf{G}^{kn}$ from Eq. (9) in terms of in-plane components $\alpha_\mathbf{G}^{kn}$ and perpendicular components $\beta_\mathbf{G}^{kn}$ as follows:

$$\mathbf{c}_\mathbf{G}^{kn} = \alpha_\mathbf{G}^{kn}\hat{z} + \beta_\mathbf{G}^{kn}\hat{\sigma}_{\mathbf{k}+\mathbf{G}}, \tag{11}$$

where $\hat{z}$ is the unit vector out of the plane and $\hat{\sigma}_{\mathbf{k}+\mathbf{G}}$ is defined by

$$\hat{\sigma}_{\mathbf{k}+\mathbf{G}} = \frac{\hat{z}\times(\mathbf{k}+\mathbf{G})}{\left|\hat{z}\times(\mathbf{k}+\mathbf{G})\right|}. \tag{12}$$

With these definitions, we define a masure of hybridization degree which will be referred to as the ellipticity because of its similarity to the common definition of ellipticity. This is defined as

$$\phi_{kn} = \begin{cases} \left( \dfrac{\sum_{\mathbf{G}} |\alpha_{\mathbf{G}}^{kn}|^2}{\sum_{\mathbf{G}} |\beta_{\mathbf{G}}^{kn}|^2} \right)^{1/2} & \text{for TM wave} \\[2ex] \left( \dfrac{\sum_{\mathbf{G}} |\alpha_{\mathbf{G}}^{kn}|^2}{\sum_{\mathbf{G}} |\beta_{\mathbf{G}}^{kn}|^2} \right)^{1/2} & \text{for TE wave} \end{cases}. \tag{13}$$

For non-magnetic problems, this reduces to $\phi_{kn} = 0$ for pure TE and TM modes. For magnetic problems, it is a natural definition of the mode mixing.

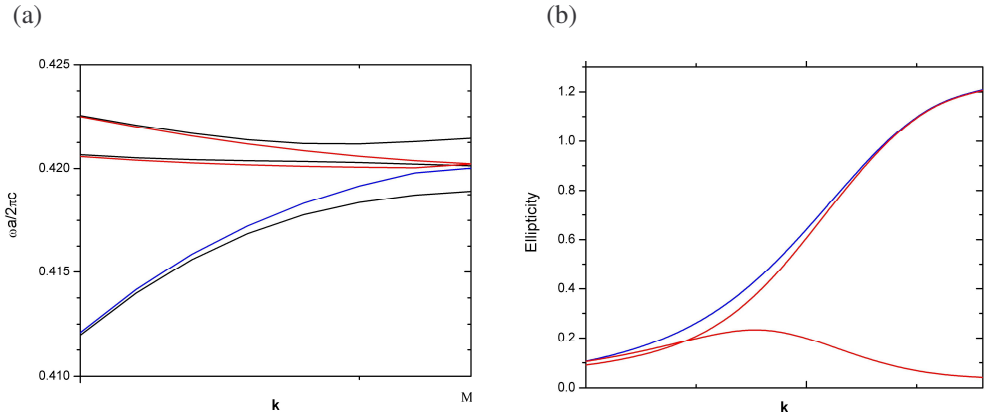

Fig. 3. (Color online) (a) Photonic band structure of MPC (shown by black lines) and non-magnetic PC (shown by blue and red lines for TE and TM polarized modes, respectively) in the proximity of M point, and (b) associated ellipticity of hybrid Bloch eigenmodes of MPC; red and blue colors are given to designate correspondence with TE and TM modes of non-magnetic structure.

In each case in Figs. 2-4, the mixing of the frequency eigenstates seen in the frequency shifts (left-hand plots) is accompanied by a hybridization of the polarization state (right hand plots). For Bloch vectors away from the degenerate regions where the frequency eigenvalues asymptote to the non-magnetic modes, the polarization hybridization tends back to the pure mode values of 0 and 1. Regardless of the nature of the degeneracy (symmetry-related or accidental), the polarization state of the Bloch mode changes from pure to hybrid.

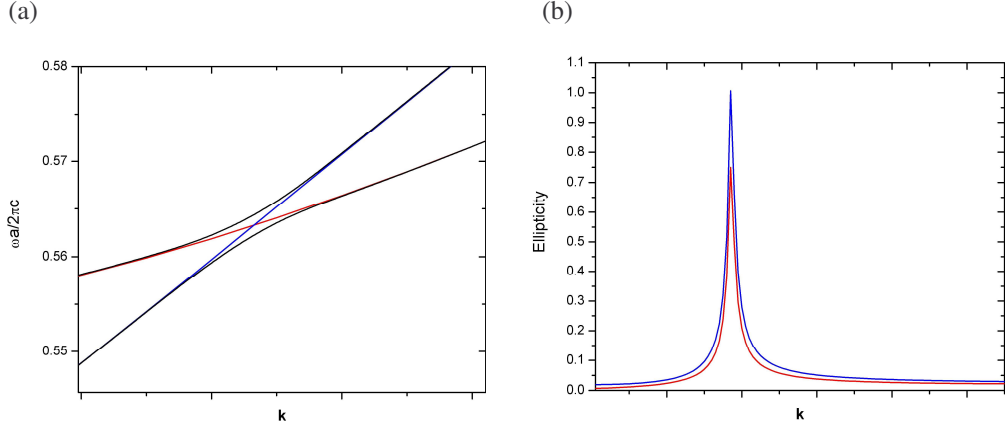

Fig. 4. (Color online) (a) Photonic band structure of MPC (shown by black lines) and non-magnetic PC (shown by blue and red lines for TE and TM polarized modes, respectively) in the proximity of accidental degeneracy, and (b) associated hybridization of Bloch eigenmodes of MPC; red and blue colors are given to designate correspondence with TE and TM modes of non-magnetic structure.

*3.1.2 Bloch mode excitation*

The coupling of the modes has interesting consequences for the excitation of the Bloch modes by light incident on the crystal from outside. Note that the MO activity couples all three degenerate and nearly degenerate eigenmodes in the proximity of the high symmetry points $\Gamma$ and M, even though they have different symmetry. In nonmagnetic structure two of these modes are symmetric with respect to the mirror reflection operation $\sigma_d$ and one mode is antisymmetric. It was shown both theoretically and in experiments [24,25] that the antisymmetric modes cannot be excited in non-magnetic PCs by an incident plane wave due to the symmetry mismatch in their field configurations.

The coupling of the modes possessing different symmetry shows that in MPCs eigenmodes cannot be classified as symmetric and antisymmetric, thus suggesting the possibility of excitation of uncoupled modes of non-magnetic PCs in their magnetic counterparts. This failure of the conventional group theory is rather expected because in the case of MPCs, magnetic group theory [26,27] should be applied instead of the conventional **k**-group method [19]. The in-plane magnetization along $\Gamma$X breaks the reflection symmetry, i.e. the reflection operation changes the magnetization direction and the transformed structure is not equivalent to the initial. It is evident however that if the magnetization direction is along $\Gamma$M, the symmetry is maintained and effect of coupling of symmetric and antisymmetric (uncoupled) modes will disappear. Calculation of PBS for MPC magnetized along the ΓM direction confirms these expectations. Coupling of the incident radiation to the modes of MPCs can be therefore dynamically tuned in magnetophotonic crystals through the control of magnetization intensity and direction.

*3.1.3 Significance of degenerate points*

Turning back to the PBS and eigenmodes we mention again that the influence of the MO activity is rather marginal far from the degeneracies. To understand this behavior, a comparison of MPCs with gyroanisotropic crystals, i.e. birefringent crystals possessing MO activity [21], is instructive [23]. In the case of gyroanisotropic crystals, birefringence significantly suppresses the effect of magnetization when the light propagation direction is not aligned with the optical axis of the crystal. This occurs because the eigenmodes of the birefringent crystals are non-degenerate linearly polarized ordinary and extraordinary waves. Due to the refractive index mismatch of the two waves, they cannot be effectively intermixed to form the circularly polarized eigenmodes inherent to isotropic MO materials [21]. In fact, the eigenmodes of gyroanisotropic crystals are in general elliptically left and right polarized modes propagating with distinct phase velocities. Thus in excluding the formation of circularly polarized eigenstates propagating with different velocities, birefringence prevents nonreciprocal polarization rotation (unless the propagation direction is chosen to be collinear with the optical axis). Now, we may consider 2D PCs as being effectively birefringent, with the TE and TM states having different effective refractive indices and playing the role of the ordinary and extraordinary waves of a birefringent crystal. However at the high symmetry points at which TE and TM modes are degenerate, they have the same effective phase refractive index. Thus propagation directions corresponding to these points play a role similar to that of the optical axis of the birefringent crystal. In the analogy with birefringent crystals, the MO activity of the MPCs in the proximity of these points results in an efficient coupling of the eigenmodes and formation of elliptically polarized modes.

Despite the similarities between MPCs and birefringent MO crystals, there are several important differences. The first is that in the case of PCs, the dispersion curves are our course folded owing to Bragg diffraction, resulting in the appearance of an infinite number of frequency dependent dispersion branches and formation of band gaps. High symmetry points, at which non-accidental degeneracies usually take place, are intersections of the Brillouin zone edges with high symmetry directions of the PC and therefore lie at the band edges. Because of this, the dispersion curves are very flat at these points, reflecting the strong Bragg scattering regime realized at the band edges with slow group velocity. As a result,

MO activity affects dispersion curves at the high symmetry points of PCs much more strongly than in homogeneous MO materials. The splitting of degenerate branches is much stronger and the difference in effective refractive indexes for hybrid TE-TM eigenmodes exceeds the value for elliptically polarized modes found in homogeneous MO materials constituting MPC. The second difference of MPCs from their homogeneous counterparts is the existence of narrow frequency ranges where only one of the hybrid modes can propagate. This implies different width of the band gaps gap for differently polarized modes. Finally, the third difference is that in PCs, the degeneracy between eigenmodes is not necessarily twofold as in the case of birefringent crystals. As can be seen from our calculations, the dispersion is three-fold degenerate at the high symmetry points of 2D MPC involving two symmetric and one antisymmetric mode (which cannot occur in homogeneous MO materials). Thus, at some conditions all three modes can be coupled giving rise to formation of magnetic Bloch waves described earlier and which cannot be classified as either symmetric or antisymmetric. These modes are formed from the nonmagnetic modes of having different symmetries.

*3.2. Two-dimensional MPCs magnetized perpendicular to plane*

For magnetization perpendicular to the plane (i.e. along $\hat{z}$), the eigenvalue problem (9) admits significant simplification as the MO activity does not introduce mixing of TE and TM modes. The problem therefore separates into two cases: (i) a vector TM problem for in-plane components of the electric field vector (or a scalar problem for the perpendicular magnetic field), and (ii) a scalar TE problem for the perpendicular electric field. While for TE polarization, the effect of MO activity is marginal (we do not consider here the trivial effects caused by uneasily magnetization induced anisotropy with axis parallel to the magnetization direction since such effects are very small of the second order in parameter |$g(r)$|), previous studies of the TM problem as well as analysis we present here have revealed several peculiarities. In fact, the TM polarization with perpendicular magnetization is perhaps the most studied geometry due to recent interest in both photonic "edge states" [15,16], and magnetic defect modes which were shown to be useful for the design of various nonreciprocal devices [11,12]. The photonic band structure for MPCs with a gyroelectric permittivity was previously studied only in Refs. [15,16]. While these works focus on the "Dirac points" of accidental band degeneracies, the authors did not analyze the polarization state of eigenmodes at these points. However, our earlier arguments indicate that these points are the portion of the Brillouin zone at which the polarization state is most significantly altered by the magnetic field.

In Fig. 5, we present calculation results for both the band structure and hybridization for this configuration. We observe that, as for the case of in-plane magnetization, the effect of MO activity is marginal for the most of the **k**-$\omega$ plane, but is significant in the proximity of the degeneracy between two branches (now both corresponding to TM-polarization in contrast to TE-TM hybridization in the parallel magnetization case). The magnetic field again induces splitting and alters modes to be in-plane elliptically polarized. (We can now speak of true ellipticity rather than polarization hybridization for the reasons explained below). Similar behavior exists in homogeneous MO materials when both the wavevector k and electric field E are perpendicular to the magnetization direction. However, the degree of ellipticity in the latter case is very small and is proportional to the gyration **g(r)** [21]. Our calculations show that in contrast to this insignificant ellipticity for homogeneous materials, in the case of MPCs ellipticity reaches unity at degeneracy points, i.e. the Bloch modes are in-plane circularly polarized waves. To confirm this unusual behavior, we calculated the Bloch modes profile within the elementary cell using the finite element method. We find that the pair of eigenmodes of the split branches possesses field profiles that rotate clockwise and counterclockwise (See. Fig. 6). At first glance, this behavior is very similar to the rotating MPC defect modes found earlier [11,12]. However, in contrast to the localized defect mode, the present Bloch modes are extended over the whole lattice of MPC and this implies that the rotating pattern is repeated in every elementary cell of the entire structure. Nevertheless, similarly to the rotating defect modes, extended rotating Bloch modes are built of degenerate modes of the nonmagnetic structure. In the case of the rotating Bloch modes, they are built of symmetric and antisymmetric TM modes of the non-magnetic PC and this fact again suggests the possibility of exciting and manipulating uncoupled modes of PCs in their magnetic counterparts.

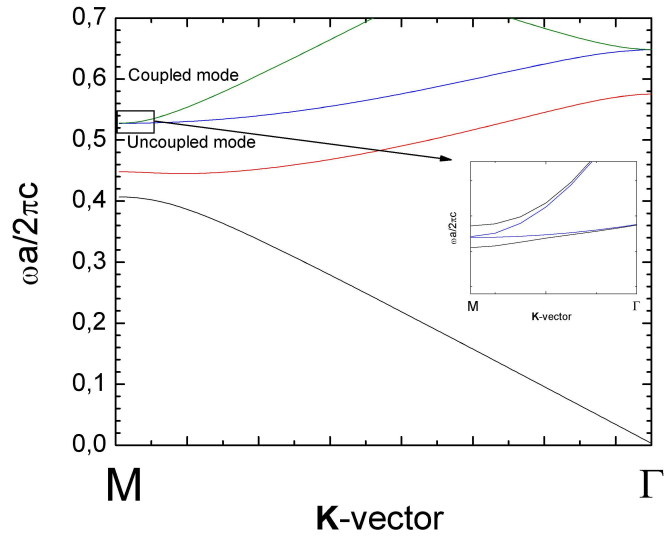

Fig. 5. (Color online) Photonic band structure and classification of bands in nonmagnetic 2D MPC. Inset shows effect of magnetization on the band structure in the proximity of M-point degeneracy: magnetic case is shown by black lines and non-magnetic by blue lines, respectively.

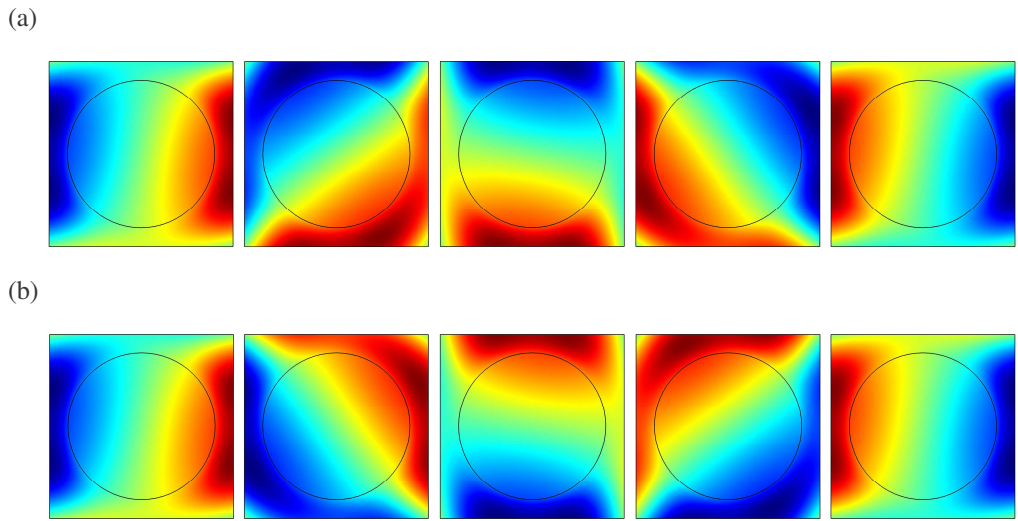

Fig. 6. (Color online) Clockwise (a) and counterclockwise (b) rotating field patterns of magnetic Bloch modes in MPC magnetized perpendicular to plane.

## 4. Three dimensional MPCs

For the sake of completeness, we now briefly discuss the effect of MO activity on the optical properties of 3D photonic crystals. This problem was thoroughly studied by Nishizawa and Nakayama [2], who studied both the alteration of photonic band structure by MO activity and the polarization state of eigenmodes. The theoretical magnetic group analysis for a 3D fcc MPC has also been presented recently in Ref. [27].

In Ref. [2], the authors predict a so-called "band Faraday effect". This effect manifests as polarization rotation for magnetization direction perpendicular to the Bloch wave-vector and its magnitude appears to scale with the square of the modulus of the gyration $|\mathbf{g}|$. In contrast, for the conventional Faraday Effect, the magnetization is collinear with the propagation direction and the rotation is proportional to the gyration. We limit our consideration to the Faraday geometry supposing the magnetization to be parallel to the Bloch vector, and consider the conventional Faraday Effect which can be interpreted as a result of the phase difference acquired by the non-degenerate left and right polarized waves.

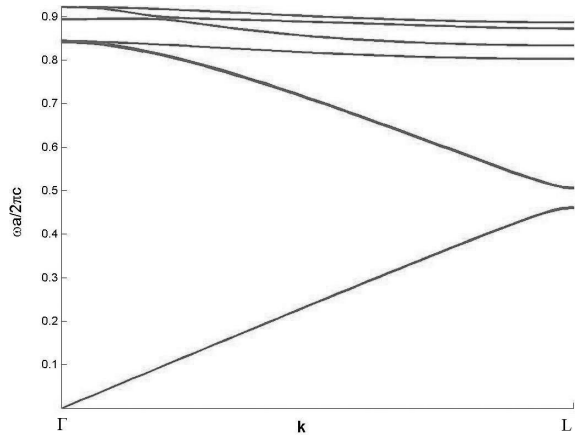

Fig. 7. Photonic band structure for 3D sc MPC composed of magneto-optically active spheres.

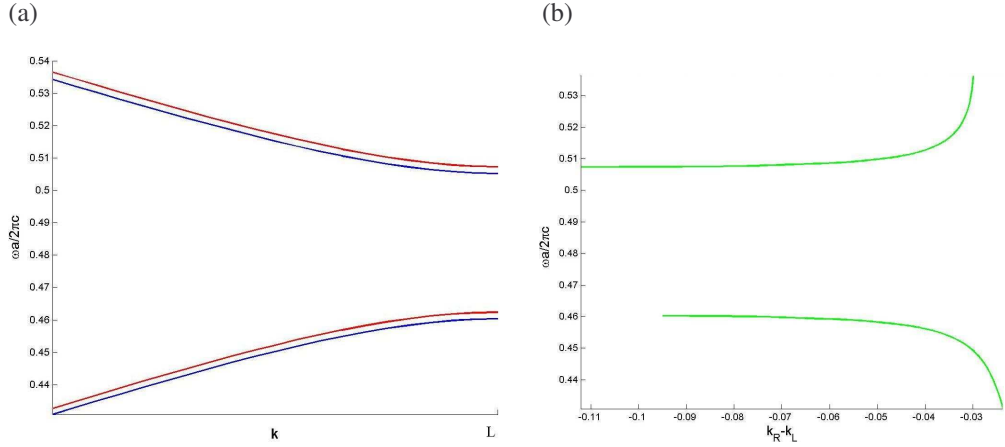

Fig. 8. (Color online) (a) Dispersion curves for the first and second circularly left and right polarized bands shown by blue and red lines, respectively, and (b) rotatory power of the 3D MPC in the proximity of the L point.

The new element in the photonic band structure of 3D PCs is the existence of high symmetry directions in k-space for which the dispersion curves for s- and p- polarized modes are degenerate, as opposed to high symmetry points in 2D. For propagation along these directions, MO activity results in the formation of genuinely circularly polarized Bloch modes so that the Faraday rotatory power can be evaluated from the difference in wave-vectors for left and right polarized eigenmodes [28,7]. For other directions, the MO activity has a limited effect, and the PBS and eigenmodes return to those of nonmagnetic PCs. Note, that for 3D structures, the similarities with gyroanisotropic crystals are even more justified, since the high symmetry directions of PCs are analogous to the crystalline axes in birefringent crystals.

Figure 7 shows the calculated photonic band structure in the ΓL direction of a 3D photonic crystal, consisting of a cubic lattice of Ce:YIG MO spheres. For this particular direction the four lower branches of spectra are doubly degenerate for s- and p-polarization. As can be seen from Fig. 8a, which shows the bands in the vicinity of the L point, the MO activity lifts this degeneracy and calculations show that the eigenmodes are circularly polarized Bloch waves. Figure 8b shows the difference between the wavevectors of these circular waves. The difference grows when approaching the band edge, implying an enhancement of the Faraday rotation due to the slow group velocity. This effect is absolutely identical to that predicted and observed in 1D MPCs [28,7], so that a significant enhancement of magneto-optical Faraday rotation is expected for band edges of 3D MPCs for light propagating along high symmetry directions [29,14].

## 5. Conclusion

The calculations presented in this work demonstrate that alteration of the PBS and eigenmodes of MPCs by magnetization is significant in the vicinity of the points where degeneracy between TE and TM polarized modes of the nonmagnetic PCs takes place. The alteration of the PBS and especially the polarization state of the eigenmodes of MPCs raises the possibility of utilizing such tunability in applications by several means: (i) manipulating the propagation direction of the light through control of dispersion, (ii) controlling the polarization states of Bloch modes, and (iii) controlling the coupling of incident radiation to magnetically tuned Bloch modes of MPCs.

In addition, the results presented here also imply that experimental study of PCs in the presence of an external magnetic field brings a convenient and simple way of determination of their photonic properties, which until now very often can be obtained only with use of complicated experimental schemes. Because of the nonreciprocity of MO media, MO effects accumulate when light is propagating in the MO material and in particular the Faraday effect is directly related to the group velocity and the density of photonic states [7]. This implies that Faraday rotation can be considered as a measure of the time spent by an electromagnetic wave within the structure, and therefore as a measure of the density of states. So measuring the MO response provides a new means for indirect determination of the group velocity and density of states. Moreover, because of the strong sensitivity of the MO effects on the direction of the propagation in MPCs, MO measurements can provide precise information about the spatial orientation of high symmetry directions inside a crystal. From this discussion it also follows that other studies of the PCs where MO effects can also be very useful are those related to disorder. In this case, comparison of the calculated and measured values of the Faraday rotation/group velocity can provide information about disorder, while orientation of the high symmetry axes will uncover distortions of the crystalline structure.

**Acknowledgments**

This work was supported in part by the Super Optical Information Memory Project from the Ministry of Education, Culture, Sports, Science and Technology of Japan (MEXT), and Grant-in-Aid for Scientific Research (S) No. 17106004 from Japan Society for the Promotion of Science (JSPS). CUDOS is an Australian Research Council Centre of Excellence.

**References**

[1] M. Inoue et al., J. Physics. D: Applied Physics **39**, R151 (2006).

[2] H. Nishizawa and T. Nakayama, J. Phys. Soc. Japan **66**, 613 (1997).

[3] A. Figotin, Y. A. Godin, and I. Vitebsky, Phys. Rev. B **57**, 2841 (1998).

[4] K. Busch and S. John, Phys. Rev. Lett. **83**, 967 (1999).

[5] A. B. Khanikaev, A. V. Baryshev, M. Inoue, A. B. Granovsky, and A. P. Vinogradov, Phys. Rev. B **72**, 03512301 (2005).

[6] A. B. Khanikaev, M. Inoue, and A. B. Granovsky, J. Magn. Mag. Materials **300**, 104 (2006).

[7] A. B. Khanikaev, A. B. Baryshev, P. B. Lim, H. Uchida, M. Inoue, A. G. Zhdanov, A. A. Fedyanin, A. I. Maydykovskiy, and O. A. Aktsipetrov, Phys. Rev. B 78, 19310201 (2008).

[8] M. Inoue, K. Arai, T. Fujii and M. Abe, J. Appl. Phys. 83, 6768 (1998).

[9] M. Inoue, K. Arai, T. Fujii and M. Abe, J. Appl. Phys. 85, 5768 (1999).

[10] M. J. Steel, M. Levy and R.M. Osgood, IEEE Photon. Technol. Lett. **12**, 1171 (2000).

[11] Z. Wang and S. Fan, Appl. Phys. B: Lasers Opt. **81**, 369 (2005).

[12] Z. Wang and S. Fan, Optics Letters **30**, 1989 (2005).

[13] A. K. Zvezdin and V. I. Belotelov, Eur. Phys. J. B **37**, 479 (2004).

[14] V. I. Belotelov and A. K. Zvezdin, JOSA B **22**, 286-292 (2005).

[15] F. D. Haldane and S. Raghu, Phys. Rev. Lett. **100**, 01390401 (2008).

[16] S. Raghu and F. D. Haldane, Phys. Rev. A **78**, 03383401 (2008).

[17] Z. Wang, Y. D. Chong, J. D. Joannopoulos, and M. Soljačić, Phys. Rev. Lett. **100**, 01390501 (2008).

[18] Z. Yu, G. Veronis, Z. Wang, and S. Fan, Phys. Rev. Lett. **100**, 02390201 (2008).

[19] K. Sakoda, "Optical Properties of Photonic Crystals," (Springer-Verlag, Berlin Heidelberg, 2001).

[20] E. Istrate, A. A. Green, and E. H. Sargent, Phys. Rev. B **77**, 1951221 (2005).

[21] A. K. Zvezdin, V. A. Kotov, "Modern Magnetooptics and Magnetooptical Materials," (Institute of Physics Pub., Bristol, Philadelphia, PA, 1997).

[22] A. Saad, "Numerical methods for large eigenvalue problems," (Manchester University Press: Manchester, 1991).

[23] A. M. Merzlikin, A. P. Vinogradov, M. Inoue, A. B. Khanikaev, and A. B. Granovsky, Journal of Magnetism and Magnetic Materials 300, 108 (2006).

[24] K. Sakoda, Phys. Rev. B **52**, 7982 (1995).

[25] W. M. Robertson, G. Arjavalingam, R. D. Meade, K. D. Brommer, A. M. Rappe, and J. D. Joannopoulos, Phys. Rev. Lett. **68**, 2023 (1992).

[26] V. Dmitriev, Eur. Phys. J. Appl. Phys. **32**, 159 (2005).

[27] V. Dmitriev, Metamaterials **2**, 71 (2008).

[28] M. Levy, H. Yang, M. J. Steel, and J. Fujita, IEEE J. Lightwave Technol. **19**, 1964 (2001).

[29] C. Koerdt, G. L. J. A. Rikken, and E. P. Petrov, Appl. Phys. Lett. **82**, 1538 (2003).